\documentstyle [12pt]{article}

\def\half{\frac{1}{2}}

\def\mt{  m_{ {\scriptscriptstyle t} }     }
\def\mH{ m_{ {\scriptscriptstyle H} }     }

\def\TC{T_{\scriptscriptstyle C} }

\def\Tone{T_{\scriptscriptstyle C} }
\def\T2{T_{\scriptscriptstyle 2} }

\def\Fc{F_{\scriptscriptstyle c} }

\def\vevphi{\langle \phi \rangle }
\def\phio{ \phi_{\scriptscriptstyle 0 } }
\def\phiplus{ \phi_{ + } }
\font\newsymb=msxm10   
                       \def\gapprox{\mathop{\hbox{\newsymb \&}}}
\begin{document}
\begin{titlepage}
\begin{center}
July   1992      \hfill     OHSTPY-HEP-T-92-016 \\

\vskip 1cm
{\Large
 {\bf Collapsing Sub-Critical Bubbles }}
\footnote{ To appear in Physics Letters B}
\vskip 1.0 cm

       {
      {\bf Greg W. Anderson }\\
	\vskip .5 cm
       Physics Department, Ohio State University, 174 W. 18th Street\\
        Columbus, Ohio 43210\\  }

\vskip 2cm
\end{center}
\begin{abstract}
In the standard scenario, the electroweak phase transition
is a first order phase transition
which completes by the nucleation of critical bubbles.
Recently, there has been speculation that the standard picture of
the electroweak phase transition
is incorrect.   Instead, it has been proposed that throughout the
phase transition appreciable amounts of both broken and unbroken
phases of $SU(2)$ coexist in equilibrium.  I argue that this can not
be the case. General principles insure that
the universe will remain in  a homogenous state of unbroken $SU(2)$
until the onset of critical bubble production.

\end{abstract}
\end{titlepage}

\section{Introduction}
\indent

Spurred by the interest in electroweak baryogenesis,
a great deal of effort has been undertaken to describe and
quantify many salient aspects of the
electroweak phase transition (EWPT) in the minimal standard model
and its extensions.  It has become standard lore that the EWPT
is a first order phase transition which proceeds by the nucleation
and the subsequent growth of critical bubbles\cite{Dine,AH,Dinelinde}.
Although the phase transition is weakly first order in the
minimal standard model, the strength of the phase  transition is model
dependent, and
the EWPT is more strongly first order in very simple
extensions of the minimal standard model\cite{AH}

Recently however,  there has been speculation that the electroweak
phase transition is actually not a first order phase
transition after all\cite{KG}.
Instead, Gleiser and Kolb have suggested that  during the phase
transition the two phases of broken and unbroken $SU(2)\times U(1)$  coexist
simultaneously with equilibrium between the two phases being established and
maintained by sub-critical bubbles.  Similar arguments have been
advanced by Tetradis\cite{Nick}.
In particular, Gleiser and Kolb and Tetradis argue
that at the critical temperature $\TC$
the universe is filled with equal parts of the broken and unbroken
phases. $\TC$ is defined as the temperature of the universe when
the free energy
density of the system, plotted as a function of $\vevphi$ has two
degenerate minima.  The basic contention of these authors is that
as long as the expansion rate is slow compared to
the rates of thermal processes (Gleiser and Kolb consider
processes mediated by sub-critical bubbles), the universe will be driven into
a state equally populated by both phases.   If true, this argument would
have important ramifications for scenarios of baryogenesis which
invoke first order phase transitions.  In addition, one might wonder if
the standard picture of the universe trapped in a homogeneous
state when the temperature reaches $\TC$ is an assumption which has should
be checked case by case or if there are general dynamical and statistical
effects which guarantee this.
Below I will argue that the basic the picture advanced by Gleiser and Kolb is
in contradiction with the second law of thermodynamics.  Other criticisms
of sub-critical bubbles have also been made.\cite{Dinelinde}
Similar remarks would apply to the analysis of Tetradis and
previous studies of subcritical bubbles\cite{KGW}.
Instead, very general properties of statistical
mechanics guarantee that the equilibrium state at temperature $\TC$
is a homogeneous state.

\section{Thermal Equilibrium at the Phase Transition}
\indent

Gleiser and Kolb and Tetradis argue that at the critical temperature
$\TC$, the universe is filled with equal parts of the broken and
unbroken phases.
The assertion that both wells are equally populated would be true
for an ensemble of particles interacting with an external potential.
However, metastability in one-dimensional
mechanics is very different from  metastability in a
field theory.   The disparate nature of these two cases is qualitative
as well as quantitative.  It is instructive to contrast these two
cases to see where some  types of intuition
can lead us astray.  Let's compare the symmetric double well  in
field theory and in one dimensional mechanics.
First  consider the one dimensional mechanical example at fixed temperature.
For an ensemble of particles interacting with the
external potential given in figure 1a, the thermal equilibrium state
of the system is one where both wells are equally populated with particles.
This situation is in sharp contrast to the case in quantum field theory.
Under conditions present at the EWPT, the thermodynamic requirement
that the total entropy of the universe can  only increase is equivalent
to demanding that the free energy only decreases.  Thus,
the equilibrium state of the system minimizes the free energy.
Recall that the free energy of the system is:
\begin{equation}
F = \int d^3 x  \,\half (\nabla \phi)^2  + V(\phi,T).
\end{equation}
For convenience we can normalize the free energy density so that
$V(\phiplus) = V(\phio) = 0$.  A universe held at a temperature $\TC$
and left to equilibrate will end up either
in the homogeneous ground state
$\vevphi = \phi_{+}$  or $\vevphi = \phio = 0$.   This is
because both the gradient
term and $V(\phi,T)$ are positive and nonvanishing inside any boundary
separating the two
phases.   So a universe filled with domains
of both phases must have a larger free energy than either homogeneous
state.  According to the second law of thermodynamics,
while an individual fluctuation can
occasionally increase the systems free energy, the cumulative
statistical effect of these
fluctuations  must decreases the
free energy. Moreover, because the universe is cooling, there is no question
which state the universe occupies when the temperature
reaches $\TC$.
The state of lowest free energy at temperatures above $\TC$ is
$\vevphi = \phio$, and when the temperature reaches $\TC$ the newly degenerate,
homogeneous vacua at $\vevphi = \phi_{+}$ is
separated from the state $\vevphi = \phio$ by an infinite
barrier.\footnote{
At temperatures above $\TC$, the metastable state
which first appears at $\vevphi = \phi_{+}$ is separated from
the homogeneous state $\vevphi = \phio$ by an infinite barrier.
Any finite {\em finite} region of space containing the new phase is not a
meta-stable state since it can be continuously deformed to the
ground state without surmounting an energy barrier.  At temperatures equal to
and
above $\TC$ finite regions of $\vevphi = \phiplus$  are completely unstable.
Only when the
temperature drops below $\TC$, the can system can be continuously deformed from
the state $\vevphi = \phio$ to the new equilibrium state
$\vevphi = \phiplus$ by crossing a finite
barrier (see figure 2).
The height of this free energy barrier is the critical bubble
free energy.}
So when the temperature drops to $\TC$
the universe finds itself in the homogeneous vacuum $\vevphi = \phio$.
This is in direct conflict with the analyses of Gleiser and Kolb
and Tetradis.

Although it is clear from these general grounds that the universe
is filled with the homogeneous state $\vevphi = 0$
when the temperature cools to $\TC$ ,
it is useful to  see the how this arises from the dynamical
equations governing the evolution of the scalar field.
This will allow us to quantify how large
fluctuations are about the equilibrium state.
Before discussing the statistical evolution of  the scalar field
it will be useful to
recall a few basic properties of nucleated bubbles.
Consider a bubble containing $\vevphi = \phiplus$  in a sea of vacuum
$\vevphi = 0$ (see figure 2).
By convention we will choose the state $\vevphi = 0$ to have
zero free energy.
Then the surplus free energy of a nucleated bubble is
\begin{equation}
F = \int d^{3}x
\left\{
\half\left( \vec{ \nabla } \phi \right)^{2}
+ V(\phi,T)  \right\}.
\end{equation}
The free energy of this bubble has two contributions:
a surface free energy $F_{S}$, coming mostly from the derivative terms
in Eq. (2.2),
and a volume term $F_{V}$, which arises  from the difference in
free energy density inside and outside the bubble.
These two contributions
scale like
\begin{eqnarray}
 F & = & F_{S} + F_{V} \nonumber \\  & \sim & \,2\pi R^{2}
\left( \frac{\delta \phi}{\delta R}\right)^{2} \delta R +
 \frac{4\pi}{3}\,R^{3} \,\bar{V}(\phiplus) ,
\end{eqnarray}
where $R$ is the radius of the bubble, $\delta R$ is the thickness of
the bubble wall, $\delta \phi \sim \phiplus $,
and $\bar{V}(\phiplus)$ is the average value of the potential inside
the bubble.
 For the bubbles we are interested in, it is energetically
favorable to make the gradient term as small as possible, so the bubbles
will be thick walled.
 For thick walled bubbles
$\delta R \sim R$, and the surface energy of the
bubble grows like $R$. In contrast, the volume term increases in
magnitude like $R^{3}$.
For temperatures below $\TC$ the volume term in equation Eq. (2.3) can
be negative (See figure 2).  At this temperature,
although the homogeneous state $\vevphi = \phiplus$
has a lower free energy, a thermal fluctuation
producing a bubble
of true vacuum which starts from a  radius of zero and expands in radius
to envelope the system, must have a free energy greater than
or equal to some critical value.  The radius of
the {\em critical} bubble
is found by  differentiating Eq. (2.3),
 $R_{c}\sim \phiplus/\sqrt{-2\bar{V}(\phiplus)}$.
Sub-critical bubbles, those bubbles with radii smaller than this
critical size,  will collapse
under their surface tension. The free energy of a
critical bubble is:
\begin{equation}
F_{c} \sim
\frac{ \phiplus^{3}}{\sqrt{-\bar{V}(\phiplus)}}.\\
\end{equation}
Notice as the temperature approaches the
critical temperature from below $\bar{V}\rightarrow 0$, and both the
radius and free energy of the critical bubble become infinite.  So
at $T\geq \TC$, all bubbles are subcritical.

   Aided by this qualitative understanding of nucleated bubbles
we can examine the dynamical equations describing the abundance of
regions containing the $SU(2)$ broken phase.
In a hot universe in the vacuum state $\vevphi =\phio = 0$,
thermal fluctuations will produce bubbles inside of which
$\vevphi$ is nonvanishing. At temperatures below $\TC$, these
thermal fluctuations produce critical bubbles at a rate per unit volume:
\begin{equation}
\Gamma /V \simeq T^4 e^{-\beta \Fc}.
\end{equation}
The exponential suppression $\exp(-\beta \Fc)$ is the
usual Boltzmann suppression
for producing configurations close to the
critical bubble, while the prefactor gives the rate of typical
processes which are not Boltzmann suppressed.
The suppression in the rate for thermal vacuum change events
arises because the system must cross
a barrier in order to produce bubbles large enough to grow.

Consider the rate at which regions of $\phiplus$ are populated
at temperatures near $\TC$.
Let $f$ denote the fraction of space filled regions of $\vevphi \sim
\phi_{+}$.  The master equation for the evolution of $f$ is:
\begin{equation}
\frac{df}{dt} = \left(1 - f\right)\Gamma(\phio\rightarrow\phiplus)
 - f \Gamma(\phiplus \rightarrow \phio)
\end{equation}
 Although the rate given in Eq. (2.5) has strictly only been motivated for
critical bubble production,
it is not unreasonable to assume that its generalization gives a
good estimate of the rate other configurations are produced.
Any fluctuation producing a region of $\vevphi \sim \phiplus$,
will be Boltzmann suppressed because energy is required to form
 the domain boundaries. So the regions of $\vevphi \sim \phio$
with spatial extent $R$ will be converted to regions of
 $\vevphi \sim \phiplus$ at a rate:
 \begin{equation}
\Gamma(\phio \rightarrow \phiplus) \simeq T(RT)^3  e^{-\beta F(R)},
\end{equation}
where $F(R)$ is the free energy of a subcritical bubble of radius $R$.
{}From Eq. (2.3), $F \gapprox 2\pi \phiplus^2 R$.
Fluctuations which create energetically
disfavored structures can also remove them.
Even in a universe filled equally with domains of both phases,
there will be fluctuations which decrease the abundance
of domain walls and take the universe toward
a homogeneous state.  Fluctuations of this sort, which decrease
the volume occupied by domain boundaries, do not cost energy
so their rates are not Boltzmann suppressed.
If anything they should be enhanced relative to rates which leave
$\vevphi$ unchanged.
Regions of $\vevphi \sim \phiplus$ with spatial extent $R$
are depleted by both fluctuations, and the dynamical collapse resulting
from the region's surface tension:
\begin{equation}
\Gamma(\phiplus \rightarrow \phio) \gapprox T(RT)^3 + 1/\tau .
\end{equation}
A simple estimate of the collapse time gives $\tau \sim R$.
Although the self induced collapse is typically  faster than
the rate of bubble production, fluctuations are even more
effective at removing regions of the unstable phase.
In steady state, detailed balance requires that the fraction of space
containing bubbles of broken phase is exponentially suppressed.
{}From Eqs. (2.6) - (2.8),
\begin{equation}
f  \leq  \frac{ e^{-F/T} }{1 + e^{-F/T} }.
\end{equation}
where $F$ is the free energy of a sub-critical bubble including the
bubble walls.
Since we are interested in bubbles which change the value of the
scalar field condensate we can set a lower limit on the
magnitude of a bubbles free energy.
In order to produce a classical shift in the scalar field
condensate, a bubble
of scalar field must contain many quanta\cite{AHH}.
Since the wavelength of a typical quantum comprising a bubble is
order $R$, with $F \sim n_{q}/R$ and $n_{q}>>1$ we must have
$(F/T) (RT)>>1$, where $n_{q}$ is the number of quanta.  Using Eq. (2.3),
\begin{equation}
F/T  \gapprox 2\pi \left(\frac{\phiplus}{T} \right)^2 (RT) >> 1/RT.
\end{equation}
In the standard model, the ratio $F/T$ satisfying Eq. (2.10) is
not small (See Figure 3).
\footnote{
In simple extension of the standard model, virtual effects of
additional particles  can make
 the phase transition proceed  as it would if
the Higgs Boson mass was significantly smaller\cite{AH}.  This
is precisely the modification which is necessary to avoid washout
of the baryon asymmetry after the phase transition has completed.
 For this reason
the graphs in figure 3 has been extended below the current
experimental limit on the Higgs Boson mass.
$\phiplus /T$ can be simply estimated using the thin wall approximation.
Note that for $\mH =
30 \,(60)$ GeV, the thin wall approximation determines $\phiplus/T$
to an accuracy of $6.5 \,\,(2.5)$ percent.}
Thus, until the onset of critical bubble nucleation,
 the universe finds itself in a homogeneous state with
an exponentially suppressed number of ephemeral regions
containing  the $SU(2)$ broken phase.

Although Eq. (2.9) demonstrates that the unbroken phase
is always favored before the phase
transition occurs, one might wonder if there are models where Eq. (2.9)
allows for a departure from the standard formalism of false vacuum
decay.  In models where $\phiplus/T <<1$ it is possible to have
$FR>>1$ with $F/T<<1$.  Indeed, we know  in the limit $\phiplus
\rightarrow 0$ we must have $f \rightarrow \half$.  However, even
in the extreme case $\phiplus/T <<1$ the standard formalism
of first order phase transitions should remain valid.  The new ground state
$\vevphi = \phiplus$ will not dominate until fluctuations can
produce regions large enough to grow, and this will not happen until
the temperature drops below $\TC$.  Whether such a region is produced
all at once or by the coalescence of smaller regions, the rate for
producing the saddle point solution is  given by Eq. (2.5).
\footnote{ When calculating the thermodynamic probability of producing a
critical bubble by a saddle point evaluation of the partition
function no choice is made to include some histories at the expense of
 others. }
The first order phase transition will occur at a temperature where
fluctuations produce regions of the new phase  large enough
to grow at a rate which exceeds the expansion rate of the universe.

\newpage

\begin{description}
\item[Figure 1:]  The symmetric double well in a mechanical
example  and in field theory.
\vskip 2cm
\item[Figure 2:]  The effective potential at temperatures near $\TC$

\item[ Figure 3:]  $\phiplus$ at the end of the phase transition
verses the Higgs boson mass for $\mt = 120$.   The dashed curve
is the thin wall approximation, while the solid curve is the
is the numerical result.  The upper and lower dotted curves
correspond to the values of $\phiplus/T$ at temperatures $\T2$ and $\Tone$
\end{description}


\newpage
\begin{thebibliography}{99}
\bibitem{Dine} M. Dine, P. Huet, and R. Singleton,
 {\it Nucl. Phys.}, {\bf B375},625 (1992)
\bibitem{AH} G. Anderson, and L. Hall {\it Phys. Rev.},
{\bf D45}, 2685 (1992)
\bibitem{Dinelinde} M. Dine, R. G. Leigh, P. Huet, A. Linde and D. Linde,
Slac preprint SLAC-PUB-5741
\bibitem{KG} M. Gleiser, and E. W. Kolb, Fermilab preprint
FERMILAB-PUB-91/305-A
\bibitem{Nick} N. Tetradis, DESY preprint DESY-91-151
\bibitem{KGW} M. Gleiser, E. W. Kolb, and R. Watkins, {\it Nucl. Phys.},
{\bf B364}, 441 (1991)
\bibitem{AHH} G. Anderson, L. Hall and S. Hsu, {\it Phys. Lett.},
{\bf B249}, 505 (1990)
\bibitem{Sher} M. Sher, Phys. Rep. 179 (1989) 275
\end{thebibliography}
\end{document}